\documentstyle[12pt]{article}

\begin{document}
\title{Confinement of Higgs bosons from hidden symmetry and convexity}
\author{Jifeng Yang$^*$ \\
\it Department of Physics \\
\it and\\
\it School of Management \\
\it Fudan University, Shanghai, 200433, P. R. China}
\footnotetext{$^*$Email:jfyang@fudan.edu.cn}
\maketitle
\begin{abstract}
In the full effective potential (EFP) approach, we find that
spontaneous symmetry breaking (SSB) can, with the convexity
of the full EFP, lead to the prediction of the IR confinement
of Higgs particles, adding difficulty to the experimental
identification of Higgs particles. The short distance behaviors
remain intact. The whole presentation is given in the
understanding that any QFT becomes UV well defined in a more
complete underyling theory according to a recently proposed
strategy.
\end{abstract}

PACS number(s): 14.80.Bn;14.80.Cp;12.90.+b
\newpage
\section {Introduction}

The standard model of particle physics within the
contemporary experiments' range has been deemed as well
established. Precision tests of standard model
have been the main activities of the particle physicists
\cite {Alt}. However, there are still some issues remaining
unclear in the standard model, the most important one is
the electroweak symmetry breaking and the Higgs
mechanism--an aspect that is still mysterious and less
understood. Many variants and/or improvements with respect
to this vital sector of standard model have been proposed and
discussed, among them there are two basic types, one in terms
of elementary scalar fields or one in terms of composite Higgs
fields. Recently some authors even try the alternative
electroweak symmetry breaking mechanism following the tricks
derived from the SUSY and string studies \cite {Delg}.

In this report we wish to study the spontaneous symmetry
breaking (SSB) phenomenon within the conventional quantum field
theory frameworks together with a new strategy to deal with
the possible UV unphysical infinities. We also wish to call
attention to the utility of the (exact) effective potential
(EFP) of a QFT via the demonstration of its role in  SSB.
This article is an extended report of the recent suggestion by
the author \cite {Yang-Higg} where we made crucial use of the
convexity of the full EFP \cite {Syman} as well as the fact
of SSB \cite {EFPSSB}.

The article is organized as follows. We discuss some 
properties and disputes about the EFP method in section II, 
especially the convexity and utility of the full EFP and
effects of renormalization. We will give a descussion basing
on our recently proposed strategy for calculating quantum 
contributions. Section III is devoted to the derivation of
the properties of the Higgs system in terms of the full EFP.
We only assume SSB mechanism and the convexity for the EFP
of the Higgs system without specifying spacetime dimension and
the other physical sectors of the system. Thus our conclusions
will not be limited to specific models. The main technical
observations are given in section III. The other implications
and discussions will also be given there. In section
IVwe make a digression on the triviality problem. The last
section will be a brief summary.
       
\section {Convexity of EFP}

First we need to discuss some properties of the main tool we
are going to use, i.e., the EFP that is also known as the
generating functional for one-particle-irreducible(1PI) Green
functions \cite {Jack} for constant field configurations
or {\sl in infrared limit}.

It is known that the full EFP is real and convex for any QFT
within which it can be consistently defined \cite {Syman}.
However, due to UV divergences, one might wonder if
renormalization could violate the convexity and there has been
a lot of literature investigating this impact
\cite {Renvex,Sher}. We will follow the standard point of view
that renormalization would not affect the convexity provided
it is appropriately done \cite {Renvex}. In fact, as will be
demonstrated shortly, if one adopts a natural strategy to deal
with the UV infinities \cite {Yang}, the convexity would
naturally follow.

In practical calculations people often arrive at nonconvex
and complex effective potentials, which seems to be in
conflict with the above assertion. The solution lies in that
the full EFP is complex where it is nonconvex \cite {Eric,Sher}.
The imaginary part arises if one starts from perturbative
definition of the EFP where the parameters (masses, couplings,
etc.) are defined in the contex of a nonconvex local
lagrangian (not an IR object), i.e., defined in a formulation
where the field configurations (or states) are not all stable
ones. It is shown by Weinberg and Wu \cite {Eric} that this
imaginary part multiplied with the space volume is half the
decay constant of the unstable modes. Orthodoxically the
arguments of EFP are not expectation values in localized
(homogeneous) states but that of superposition of distinct
states in the nonconvex region \cite {Eric,Sher,Curt}.

In this report, however, we follow a somewhat unorthodox line
of argument. It is known that the Landau-Ginsurg model is a 
phenomenological and nonconvex model \cite {Huang}--merely a
simple polynomial potential in terms of a few
phenomenologically-defined coefficients. There are some
ingredients in this model that are not thermodynamically
stable. Similar thing happens in the van der Waals theory.
Now we ask the following question: what will the complete
and thermodynamically stable formulation look like? The
answer will be a consistent formulation with all
thermodynamically unstable states removed and convex
thermodynamic potentials \cite {RU} will naturally follow
from the very thermodynamical stability. In the quantum
field theory with a nonconvex phenomenological lagrangian,
one would naturally ask: {\bf what are the final outcome
of the decay indicated by the imaginary part?} For the
decay to stop with the imaginary part gone, the final
outcome will necessarily be a formulation containing only
stable modes or states, then there should be no imaginary
and/or nonconvex parts for the full EFP at all and no
general obstacle to interpret the EFP as mininal energy
functional for homogeneous field expectation values as
IR limit of the full effective action.

There remains one issue to be addressed, the meaning of any
possible flat bottom of the EFP. Since the full EFP 
should have taken in the all quantum effects, this region is
effectively isolated in the large distance from all
the other part where the EFP is not flat (Cf. section III).
This is an alternative way of securing the stability of the
physical vacuum. Conventionally, one would resort to a
nonconvex lagrangian and define the qunatum theory around
a local minimum to stablize the theory. But an imaginary
part would necessary appear. So unlike the orthodoxical
point of view that the convex EFP is not quite useful,
our unorthodox arguments shows that the convex EFP is
informative provided it is understood as defined in the
context without unstable modes. More conservatively,
we are trying an interpretation of the flat bottem so that
the convex EFP might still be physically useful.

Hence, to get a real and convex full EFP, the qunatum theory
should be formulated in terms of stable field configurations
and states. (Here we emphasize that this convexity of the
full EFP does not necessarily mean that the short distance
structures of the theory should also be simply described by
a convex microscopic model as the EFP is only a meaning full
object in the IR limit.) It is a demanding job to find such
a formulation, which might be a very complicated quantum
theory. For our purpose here, we only need to assume that such
a formulation exists (the SSB (scalar) sector will still be
named as the Higgs sector but without the nonconvexity
in the original Higgs model\cite {Higg}).

Now we serve the simple proof of the convexity of EFP
following our strategy for calculating the quantum
corrections proposed by the author \cite {Yang}. In our
proposal, we can view a conventional QFT of standard model
as an ill-defined (due to UV and/or IR divergences)
effective sector of a complete and well-defined underlying
theory. In the underlying theory the path integral of the
QFT should read (to shorten the presentation we only write
out the SSB sector with the other parts either integrated
out or kept external for appropriate  purposes)
\begin{equation}
Z_{\{\sigma\}}(J) =\int D\mu(\phi_{\{\sigma\}})
exp[i\{S_{\{\sigma\}}(\phi_{\{\sigma\}})+
(\int J*\phi_{\{\sigma\}})\}] 
\end{equation}
where $J$ denotes the external sources (spacetime dependence
is understood here), the subscript
$\{\sigma\}$ refers to the underlying structures and it is
used to indicate that Higgs fields and the action are effective
objects in the sense of the underlying theory. This functional
is by postulate well defined and the path integration can be
done. Since the underlying structures are infinite comparing
to the effective objects, a limit operation is triggered on
the functional and therefore the functional would be given in
term of the low energy parameters and some possible finite
constants (arising from the limit operation) in place of
divergences. It is illegitimate to simply perform the limit
operation first since the path integration and limit operation
do not always commute. Otherwise, divergences would show up.
With this observation, a simple strategy for calculating loop
corrections without knowing the exact underlying structures
follows naturally and it applies to any physically meaningful
model (renormalizable or not), see Ref. \cite {Yang}. Note
again, it is understood that the effective formulation thus
obtained is free of any unstable ingredients.

Due to the arguments given above, we have 
\begin{equation}
Z(J;\bar c ) \equiv exp[iW(J;\bar c )]
=L_{\{\sigma\}} Z_{\{\sigma\}} (J) \equiv
 L_{\{\sigma\}} exp[iW_{\{\sigma\}}(J)],
\end{equation}
where $W$ refers to the connected functional and $\bar c$ are
the finite constants arising from the limit operation. 
In conventional renormalization programs, one introduces some
regularization acting as artificial substitute for the true
underlying strutures and later performs subtractions to
remove the artificial structures. However, in our point of
view one must be careful about this artificiality and
subtractions as is evident from the explicit existence of the
constants $\{\bar c\}$ in Eq.(2). Without knowing the
underlying structures we have to resort to appropriate physical
principles and experimental facts for relevant physics for
defining these constants. Here it suffice to assume that we
can define the constants in the way equivalent to the
underlying theory defintion.

Now let us perform the Legendre transform on the connected
functionals to find the effective actions or the generating
functionals for 1PI Green functions. Then there are two such
generating functionals, one with underlying parameters and
one with the constants $\bar c$ instead:
\begin{eqnarray}
\Gamma_{\{\sigma\}} (\phi_{\{\sigma\}})
&\equiv& LgT [W_{\{\sigma\}}(J)] \equiv
        Min_J\{\int \phi_{\{\sigma\}} * J-W_{\{\sigma\}}(J)\}
       \nonumber \\
&=& \int \phi_{\{\sigma\}} * J^0-W_{\{\sigma\}}(J^0), \ \ \ \
         \phi_{\{\sigma\}} \equiv 
         \frac{\delta W_{\{\sigma\}}(J^0)}{\delta J^0};\\
\Gamma(\phi ;\bar c )
&\equiv& LgT [W (J;\bar c)] \equiv
        Min_J\{\int \phi *J-W(J;\bar C)\} \nonumber \\
&=&\int \phi * J^0-W (J^0;\bar c), \ \ \ \ \ 
        \phi \equiv \frac{\delta W(J^0;\bar c)}{\delta J^0}.
\end{eqnarray}
Note that by definition the effective action is defined at
a functional point $J^0$ that maximizes $W$.
Since the Legendre transform $LgT$ and the limit operation
$L_{\{\sigma\}}$ apply to different arguments and all the
objects here are well defined, we have
\begin{eqnarray}
& & [LgT,L_{\{\sigma\}}] = 0, \\
& &\Gamma(\phi ;\bar c)
\equiv L_{\{\sigma\}}\Gamma_{\{\sigma\}}(\phi_{\{\sigma\}}),
   \ \ \  \phi \equiv L_{\{\sigma\}} \phi_{\{\sigma\}}.
\end{eqnarray}

Now it is straightforward to see that the effective action
is a concave functional due to definition in Eq.(3), i.e.,
\begin{equation}
\frac {\delta^2\Gamma_{\{\sigma\}}}
      {\delta\phi_{\sigma}(x)\delta\phi_{\sigma}(y)}
      =\frac {\delta J^0_x}{\delta\phi_{\sigma}(y)}=
\{\frac {\delta^2W_{\{\sigma\}}(J^0)}
  {\delta J^0_x\delta J^0_y}\}^{-1}
   \le 0.
\end{equation}
Note that the nonpositive second order functional derivative
only symbolically inidcates that $J^0 (x)$ maximizes $W$.
Applying the limit operation to Eq.(7) would lead to the same
conclusion for $\Gamma(\phi;\bar c)$. One can also use Eq.(4)
to prove its concavity (in the functional sense):
\begin{equation}
\frac {\delta^2 \Gamma(\phi ;\bar c)}
  {\delta\phi_x\delta\phi_y}
=\frac {\delta J^0_x}{\delta \phi_y}  =
\{\frac {\delta^2W(J^0;\bar c)}
  {\delta J^0_x\delta J^0_y}\}^{-1}
    \le 0.
\end{equation}
Again the inqeuality takes a symbolic meaning.
Then the EFP for constant $\phi$ reads
\begin{eqnarray}
U_{eff} (\phi ;\bar c)=
-\Omega \Gamma (\phi ; \bar c), \\
 \phi (x) = const.,\ \ \ \ \Omega \equiv \int d^n x
\end{eqnarray}
and the EFP's convexity follows as a consequence of the fact
that the effective action is a concave functional. In the
course of the derivation the constants $\bar c$ do not make
any trouble once they are consistently defined. As the
orthodox renormalization procedures just amount to afford
definitions for these constants, we conclude that consistent
renormalization programs would not violate the convexity
of the full EFP in renromalizable models. If one adopts
our approach, the conclusion applies to any model with
consistent definitions for $\bar c$ implementable.

In the following we will not label out the constants $\bar c$.
But the flavor degree will be labelled out where necessary.

\section {Hidden symmetry and confinement of Higgs fields}

Now let us write down the convexity relation as
\begin{equation}
\partial_{\phi_i} \partial_{\phi_j} U_{eff} ({\bf \phi})
   \geq 0, i,j=1, \cdots, N
\end{equation}
with the N-Flavor index denoted by $i,j$.

Then the SSB on flavor symmetry can be stated as:
$U_{eff}({\bf \phi})$ is invariant under the action
of the symmetry group $G_{flavor}$ while the vacuum state
$|0 \rangle$ is not, that is to say
\begin{equation}
\hat U (g) |0 \rangle \neq |0 \rangle\>, g \in G_{flavor}
\end{equation}
with $\hat U (g)$ denoting the unitary representation of the
group $G_{flavor}$ in QFT and 
${\bf \phi}_{vac}=
\langle 0|\hat {\bf \phi}|0 \rangle (\neq 0)$
minimizes $U_{eff} ({\bf \phi})$:
\begin{equation}
\partial_{\phi_i} U_{eff} ({\bf \phi}) =0, \ \
\partial_{\phi_i} \partial_{\phi_j} 
U_{eff} ({\bf \phi}) \geq 0
\end{equation}
in a small neighborhood of ${\bf \phi}_{vac}$, or equivalently
\begin{equation}
U_{eff} ({\bf \phi}) \geq U_{eff} ({\bf \phi}_{vac}), \ \ \ \
 \forall {\bf \phi}
\end{equation}
while the degeneracy of the vacuum state indicates the
existence of Goldstone modes \cite {Gold}.

Now combining Eqs. (13),(14) and (11), it is very easy to
conclude that the EFP must have a flat bottom, i.e.,
\begin{equation}
U_{eff} ({\bf \phi}) \equiv U_{eff} ({\bf \phi}_{vac}), \ \ 
\forall {\bf \phi} \in {\bar A}:=\{{\bf \phi}: 
|{\bf \phi}|\leq |{\bf \phi}_{vac}|\}.
\end{equation}
Then obviously,
\begin{eqnarray}
\Gamma^{(n)}_{i_1 \cdots i_n}(0,0,\cdots,0):=
\partial_{\phi_{i_1}} \cdots \partial_{\phi_{i_n}} 
U_{eff} ({\bf \phi}) \equiv 0 ,\ \ \ \forall n \geq 1,
 \\ \nonumber
{\bf \phi}\in A^0 := \{{\bf \phi}: |{\bf \phi}|
 < |{\bf \phi}_{vac}|\}
\end{eqnarray}
while $\Gamma^{(n)}_{i_1\cdots i_n}$ could not vanish identically
outside the set $A^0$ for at least $n=2$, which is the
two-point 1PI Green function at zero momentum---the effective
mass, then there must be at least one index $i_0$ such that
$\partial_{\phi_{i_0}}^2 U_{eff}
\left. \right |_{{\bf \phi}={\bf \phi_{vac}}}>0$
otherwise all fields would be massless ones which is
certainly not the case in SSB sector. Then we immediately
have

{\bf Proposition I}

{\it The full effective potential for the Higgs model with SSB
could not be expanded into analytical Taylor series around the
vacuum state or any state degenerate with the vacuum. In other
words, the Higgs sector (SSB sector) is singular in the IR
limit.}

Since $\Gamma_{i_1 \cdots i_n}$ assume effective interactions
{\bf in the IR limit}, it follows immediately that in the IR
limit and without gravitation the sector $\AA$ is a totally
free sector with only massless states and each state (modulo
degeneracy) in this sector is isolated with any other one
(including states beyond $A^0$) due to the absence of effecitve
interactions. Of course the standard model could not stand on
any state in this massless sector but only be established on
the physical vacuum state ${\bf \phi}_{vac}$ that could
not transit into the $\AA$ sector. This is just the mechanism
assuring the stability of the physical system realized in the
context of convex full EFP in the unorthodox reasonings
advocated in section II. Note that the IR singularity
predicted here for the SSB sector is a genuine physical
property of the theory. We may associate discontinuity in the
effective vertices with phase transition behavior, with Higgs
fields acting as order parameters. Then the phase transition
is a second order one. Thus the convex EFP (or stable
formulation) is not uninformative. It is important to recall
that the derivation here is based on the full theory, not
perturbative approximation, and hence effective vertices and
their generating functional are nonperturbative objects. 

Since Proposition I tells us that we can not Taylor expand
the full EFP around the vacuum, this immediately implies that
the effective couplings for the Higgs modes that breaks
the flavor symmetry spontaneously are {\bf infinite in the IR
limit}, i.e.,

{\bf Proposition II}

{\it In a model with SSB sector, the usual IR asymptotic
scattering states can not be defined in the full theory for
Higgs fields. These fields's quanta are subejct to infinite
effective couplings in the IR limit. In standard model, the
Higgs particles seem to be IR confined
in a formulation constructed with stable parameters only.}

This confinement predicted here does not tell us explicitly
anything about the theory's short-distance behavior since
the EFP is defined in the IR limit. It is quite natural that
the mass term and the effective couplings for the Higgs fields
remain tractable and well behaved in the short-distance
processes and all the conventional calculations and conclusions
about the short-distance behavior should remain valid. The
origin of the IR singularity of the EFP can be roughly
understood in the following way:
the effective action $\Gamma (\phi)$ for the QFT should
be well defined as a functional of the spacetime dependent
expectation value of Higgs fields ($\delta\phi (x)$) as well
as all the 1PI Green functions ($\Gamma^{(n)} (x_1,x_2,
\cdots,x_n), \forall n\geq 2$). Then the effective
vertices in EFP read
\begin{equation}
\Gamma^{(n)}(0,\cdots,0)
=\int \prod_{j=1}^n dx_j \Gamma_c^{(n)} (x_1,x_2,\cdots,x_n)
\end{equation}
where the integrations are over the whole spacetime. Then
even if the 1PI $n$-point functions given in the effective
action are well-defined functionals of $\phi(x)$, the
infinite range of spacetime integrations might give rise
to singular objects. Suppose an $n$-point function differs
a tiny amount at different points of a neighborhood of
$\phi$, then the infinite spacetime integrations will
make this difference explosively amplified, i.e., the
left hand side of Eq.(17) will become singular. In other
words, the spacetime integration could turn a regular
object into a singular one. More spacetime integrations,
more singular the effective vertices are, in
perfect accordence with the above discussions about the
vertices in EFP. Thus, in a sense, despite the IR
singularity we might expect the theory be regular at short
distance with the Taylor type functional expansion
being feasible and hence the Higgs fields may possess
scattering states in the short distance, perhaps like
quarks somehow. The short-distance behavior is dictated by
the original effective action with spacetime dependent
field configurations while the IR
behavior is dictated by the effective potential
with field configurations being constant in spacetime.
Of course the short distance should not be so short that
the standard model is invalid and new dynamics
sets in.

The mass bounds for Higgs particles, in our point of view
here, should be bounds for Higgs masses effectively
defined in the short distance. Our use of full convex EFP
(or equivalently the formulation with stable parameters only)
here seems to make us for the first time to be able to predict
the confinement (of Higgs fields) from SSB within the
conventional QFT frameworks.

One might be supprised at the prediction here of the confinement
in the Higgs model. If the prediction is valid, where does it
come from given that the Higgs couplings are believed to be not
IR singular? First we note that we did not specify the Higgs
sector in our dicsussions, only SSB is required. Since the
original Higgs model is only one way of realization of SSB in a
phenomenological sense, there is no point to extrapolate
the running coupling behavior there to all the other
formulations realizing SSB. At least we do not know the
true underlying theory yet and can not exclude the
possibility of Higgs confinement right now. Second, even
within the original Higgs model there is the nonconvex
piece which is unstable, after these unstable ingredients
decay way, the stable dynamics would not be like the
original Higgs model any more, and hence there is chance
for interactions effectively leading to Higgs confinement
or we can expect that the decay mechanism of the unstable
modes has something to do with the IR confinement of the
stable modes.

It is not clear whether the confinement indicated here is
similar to color confinment, as the dominating interactions
for the two sectors are different. It is also not clear
what the Higgs particles are confined into. We had not made
explicit dynamical calculation of the running coupling
constants here, but we found some qualitative constraints
on the effective coupling constants imposed by very general
principles of the theory--SSB and convexity (or equivalently
stable parameter formulation), a nonperturbative result.
The prediction that the IR confinement of Higgs fields
follows from the flavor symmetry breaking (SSB) complies
with the well known fact that in some abstract models Higgs
and confined phases are indistinguishable \cite {Fradk}. In
fact our arguments here add to support that relation between
Higgs and confined phases. The present investigation might
hint a new scenario for the particle physics due to the scale
differences: as the energy goes down, quarks become
confined first above $\Lambda_{QCD}$, then the Higgs
particles become confined at still lower scale. If we still
trust the Higgs model with $\lambda \phi^4$ couplings in
the short distance (with energy scale no larger than the
scale where the model fails, however) and accept the IR
property revealed in the convex EFP (unstable modes removed
via Legendre transform), then we roughly have a
weak-coupling confining phase with SSB occurring in the
meantime. Very recently a weak-coupling realization of
confinement and chiral symmetry breaking has been discussed
where confinement and SSB seem to concur, the so-called
color-flavor locking phenomenon\cite {CFlock}. This, may
serve as a third argument in favor of the Higgs confinement in
standard model in addition to the two given above.

Given the recent result of Higgs mass range ($m_{H}=
115^{+116}_{-66} GeV/c^2$, \cite {Fit}) (in the short distance
dynamics), we should
be reminded of any possible unconventional properties or
aspects of the SSB in addition to the conventional wisdoms in
the course of Higgs hunting. The confining picture for Higgs
fields here might suggest that we should pay more attention
to objects besides normal IR scattering states. Of course the
underlying microscopic dynamical mechanisms that lead to SSB
together with IR confinement of Higgs particles is still out
of our sight. As our investigation only made use of a few
general properties (mainly SSB as convexity should be a
natural property for theory with stable parameters only),
the confinement phenomenon predicted here might be a model
independent and universal one somehow. We also wish to
mention that it is believed for decades that color
confinement in QCD implies chiral symmetry breaking
\cite {Casher}, while our investigation here seems to
demonstrate a reverse situation for Higgs sector, i.e.,
symmetry breaking implies (IR) confinement. Both point to a very
close relation between SSB and confinement. So further
investigations on the subject and its relevance to the
quark confinement, especially to the confinement in the
supersymmetric gauge theories \cite {Shifman}, will be
interesting and important. Of course our prediction here
based on the EFP approach should be checked independently
in other frameworks. We would like to add that the conclusions
in SUSY QFT follow from a very simple property--holomorphy
of the Wilsonian effective action, while ours follows
from a very simple property--convexity of the full
effective potential. And our prediction here is by
now not in contradiction with the established theoretical
and experimental facts, at least in principle.

We would like to say that even
one doubts such use of the full EFP, the instability in
the nonconvex formulation indicated by Weinberg and Wu's
work \cite {Eric} suggest that quantum theory of the SSB
(or Higgs) sector might be far more complicated then
traditionally expected. There might be some important
new aspects, if not the confining picture predicted here,
to be revealed in the SSB sector of standard model.
The use of convex EFP here at least can help to remind us
of such possibility about SSB. 

\section {Digression on Triviality}

Now we would like to digress on the triviality issue for
a while which is often associated with scalar QFT
\cite {Trivia}. Most QFTs known by now are beset with certain
kind of ill-definedness somehow. They should in fact be
effective theories only valid {\sl within a finite energy
range}. There are underlying structures that, if correctly
formulated, could remove the ill-definedness in QFTs.
This is what is now accepted as standard point of view
and closely followed in our approach for renormalization
\cite {Yang}. We had pointed out that in section II,
according to the standard point of view, the present QFT
formulation amount to be resulted from an illegitimate
operation: let the underlying structures vanish before the
intermediate quantum processes are fully accounted. This
operation then calls for the need of regularizing the
theory by hand and then subtracting the divergences
afterwards.

This artificiality might make a QFT fail to describe the
physics fiathfully, since it has effectively deformed the
true underlying structures in a way unknown to us.
Such examples are not rare in literature especially
in unrenormalizable cases. Thus it is probable that a bad
regularization scheme made a theory inconsistent or
trivial, due to untamed artificiality introduced by the
regularization within the theory's validity realm.
Wonderful regularization schemes are rare to find, and
the true underlying structures remain elusive to us.
Things become worse when one tries to extend the energy
scale of the effective QFT to a place where the theory is
no longer supposed to be valid. In this case the theory, no
matter in what kind of regularization, {\sl should be
abandoned and superceded by other effective theories},
there is no point in using and discussing it any more
\cite {Yang}. This time, blames should not be put on the
theory, but on the user. Thus, it is important to probe the
boundaries where an effective theory fails and keep in mind,
when making predictions, these boundaries as well as the
influence of the artificiality residing in a regularization
scheme.

A truly trivial theory should not be able to yield any
nonzero effective interactions at any
scale, in any regularization schemes. Once a quantity in a
QFT becomes 'trivial' might imply that one
had crossed the validity boundaries of the theory except for
the possibility that the theory is truly trivial and totally
useless in the traditional sense. In fact no theory is
totally useless provided it has been constructed following
genuine physical principles and due consistency. (The
regularization artificiality should be carefully and
effectively removed already). The only problem is that
some theories constructed are less predictive or are valid
within smaller ranges.

The IR singularity predicted here might make one think
that the theory is IR trivial according to conventional
wisdom about triviality. In our point of view, since our
prediction is based on such a formulation that all the
the no unphysical divergence should show up (or a
formulation that reasonable procedures removing both UV and IR
divergences should have been done) and all the quantum
corrections accounted, i.e., a nonperturbative prediction
in terms of well-defined parameters, the singularity
can not be a signal of triviality in the IR limit. It
literally implies that the theory is physically singular
in terms of the Higgs fields, but not necessarily IR
singular in terms of other fields or parameters,
i.e., the objects into which are the Higgses confined.
Recall that in QCD, the IR singularity leads us to
conclude the color confinement instead of triviality.

\section {Summary}

Again we note that since our prediction here do not depend
on the model specifics, the
phenomenon of confinement following from symmetry breaking
might be somehow universal. In other words, the underlying
dynamics leading to SSB might also dictate the confinement
phenomenon. The close relation between quark confinement and
chiral symmetry breaking has been interesting theorists
for several decades. The prediction here, yet to be checked
independently in other approaches, might add to the ongoing
interests upon the investigations in SSB and confinement.

In short, we just suggested an alternative angle of looking
at the SSB phenomenon. Without doing detailed dynamical
calculation, we found some nontrivial consequences following
from SSB basing quite general and plausible assumptions, i.e.,
consistent existence of convex full effective potential or
equivalently the existence of a formulation of standard model
in terms of stable parameters only. The
interesting IR confinement of the fields triggering SSB
is not, at least in rough sense, in contradiction with
known theoretical and experimental facts.


\end{document}